\documentclass[twocolumn,a4,pra,superscriptaddress]{revtex4-1}
\usepackage{amsfonts}
\usepackage{setstack}
\usepackage{graphicx}
\usepackage{color}
\newcommand{\gtrsim}{>}

\newcommand{\avg}[1]{\left< #1 \right>}

\newcommand{\dintc}[1]{\mathrm{d} #1 \,}

\newcommand{\eqref}[1]{(\ref{#1})}

\newcommand{\X}{\hat{x}}
\newcommand{\Y}{\hat{y}}
\newcommand{\Z}{\hat{z}}

\newcommand{\bra}[1]{\langle #1 |}
\newcommand{\ket}[1]{|#1\rangle}

\newcommand{\hc}{\mathrm{h.c.}}

\begin{document}

\title{Majorana box qubits}

\author{Stephan Plugge}
\affiliation{Center for Quantum Devices and Station Q Copenhagen, Niels Bohr Institute, University of Copenhagen, DK-2100 Copenhagen, Denmark}
\affiliation{Institut f\"ur Theoretische Physik, Heinrich-Heine-Universit\"at, D-40225  D\"usseldorf, Germany}
\author{Asbj\o rn Rasmussen}
\affiliation{Center for Quantum Devices and Station Q Copenhagen, Niels Bohr Institute, University of Copenhagen, DK-2100 Copenhagen, Denmark}

\author{Reinhold Egger}
\affiliation{Institut f\"ur Theoretische Physik, Heinrich-Heine-Universit\"at, D-40225  D\"usseldorf, Germany}

\author{Karsten Flensberg}
\affiliation{Center for Quantum Devices and Station Q Copenhagen, Niels Bohr Institute, University of Copenhagen, DK-2100 Copenhagen, Denmark}

\begin{abstract}
Quantum information protected by the topology of the storage medium is expected to exhibit long coherence times. Another feature are topologically protected gates generated through braiding of Majorana bound states. However, braiding requires structures with branched topological segments which have inherent difficulties in the semiconductor-superconductor heterostructures now believed to host Majorana bound states. In this paper, we construct quantum bits taking advantage of the topological protection and non-local properties of Majorana bound states in a network of parallel wires, but without relying on braiding for quantum gates. The elementary unit is made from three topological wires, two wires coupled by a trivial superconductor and the third acting as an interference arm. Coulomb blockade of the combined wires spawns a fractionalized spin, non-locally addressable by quantum dots used for single-qubit readout, initialization, and manipulation. We describe how the same tools allow for measurement-based implementation of the Clifford gates, in total making the architecture universal. Proof-of-principle demonstration of topologically protected qubits using existing techniques is therefore within reach.
\end{abstract}

\maketitle

Majorana bound states (MBSs) in topological superconductors (TSs)  have been identified as promising candidates for topological carriers of quantum information \cite{Nayak2008,Alicea2012,Leijnse2012,Beenakker2013}. They may be realized as end states of proximitized semiconductor nanowires, and experimental evidence for their existence  is rapidly mounting \cite{Mourik2012,Higginbotham2015,Albrecht2016,Zhang2016}. The topological nature of MBS systems gives rise to quantum information being stored in a non-local state, not measurable by local operators. This property is intimately connected to non-Abelian braiding of MBSs, meaning that readout results depend on the order in which MBSs are brought together and measured \cite{Nayak2008,Clarke2011,Alicea2011,Hyart2013,Aasen2016}. Experimentally, however, the required branching of topological wires ($T$-junctions) is challenging. The question how topological qubits could otherwise be verified naturally arises. Here our purpose is to design circuits allowing for topologically protected storage and manipulation of quantum information in structures without branched topological segments.

\begin{figure*}
\includegraphics[width=2.1\columnwidth]{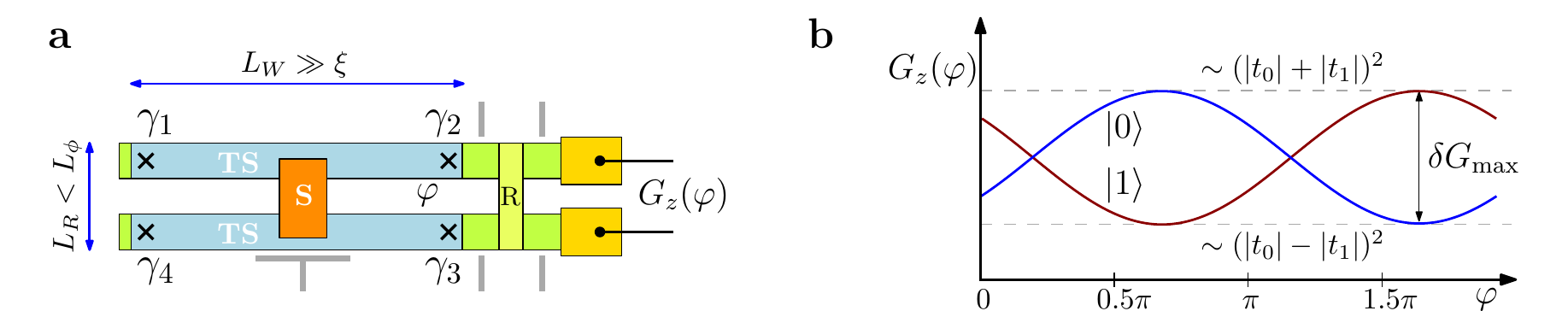}
\caption{\label{fig1}
Majorana box qubit and readout based on conductance interferometry.
{\bf a}, Two long TS wires (blue) are shunted by a superconducting bridge (S, red) to form a floating island hosting four Majoranas $\gamma_j$ (crosses).
We study long wires $L_W\gg \xi$, with $\xi$ being the TS coherence length, such that  MBS wave function overlaps are negligible and all MBSs are zero-energy states. With gate electrodes (grey), one can adjust tunnel couplings through the non-proximitized semiconductor regions (green).   The few-channel semiconducting reference arms ($R$, light green ) are shorter than their phase coherence length, $L_R < L_\phi$,  and a dimensionless magnetic flux $\varphi$ is enclosed by the resulting interference loop.
The electrostatic potential is controlled by a nearby gate and tuned to a Coulomb blockade valley with quantized charge on the island. Readout of $\Z=i\gamma_2\gamma_3$ is possible via conductance interferometry between two normal leads ({yellow}).
{\bf b}, The conductance $G_{z}(\varphi)$ is $2\pi$-periodic in $\varphi$, with a relative $\pi$-shift for the two qubit states $|0\rangle$ and $|1\rangle$ with $\Z$-eigenvalue $z=+1$ and $z=-1$, respectively.  To achieve good readout fidelity, one may tune the flux to a point of maximum contrast with $\delta G_\mathrm{max} \sim 4|t_0 t_1|$, cf.~equation~\eqref{conductance1}.}
\end{figure*}

The core of our design is the Majorana box qubit (MBQ) formed in a floating superconducting island with two long TS nanowires, marked $TS$ in Fig.~\ref{fig1}(a). This structure is experimentally attractive because the parallel wires can be driven simultaneously into the TS phase by a uniform magnetic field, and because the connecting transverse superconductor ($S$) can be a conventional $s$-wave superconductor. In addition, the designs require reference arms that can be non-proximitized semiconductors, and hence there are no $T$-junctions of topological superconducting wire segments. The wire geometry is natural for interfacing the qubit with quantum dots, employed to read out and manipulate the stored quantum information. The quantum dots can be defined by gates at the segments of the wire which are not part of the box, see Fig. 1. The parity of the MBQ is protected by its charging energy which is supposed to be large. It is important to note that the charging energy scales linearly $\sim 1/L_W$ with the size of the box, while the residual energy splittings of the Majorana modes are exponentially suppressed for increasing TS wire length $L_W$. Moreover, for the quantum dot schemes discussed below, the charge of the isolated MBQ system is fixed, thus protecting the qubit from quasiparticle poisoning.

\begin{figure*}
\centering
\includegraphics[width=2\columnwidth]{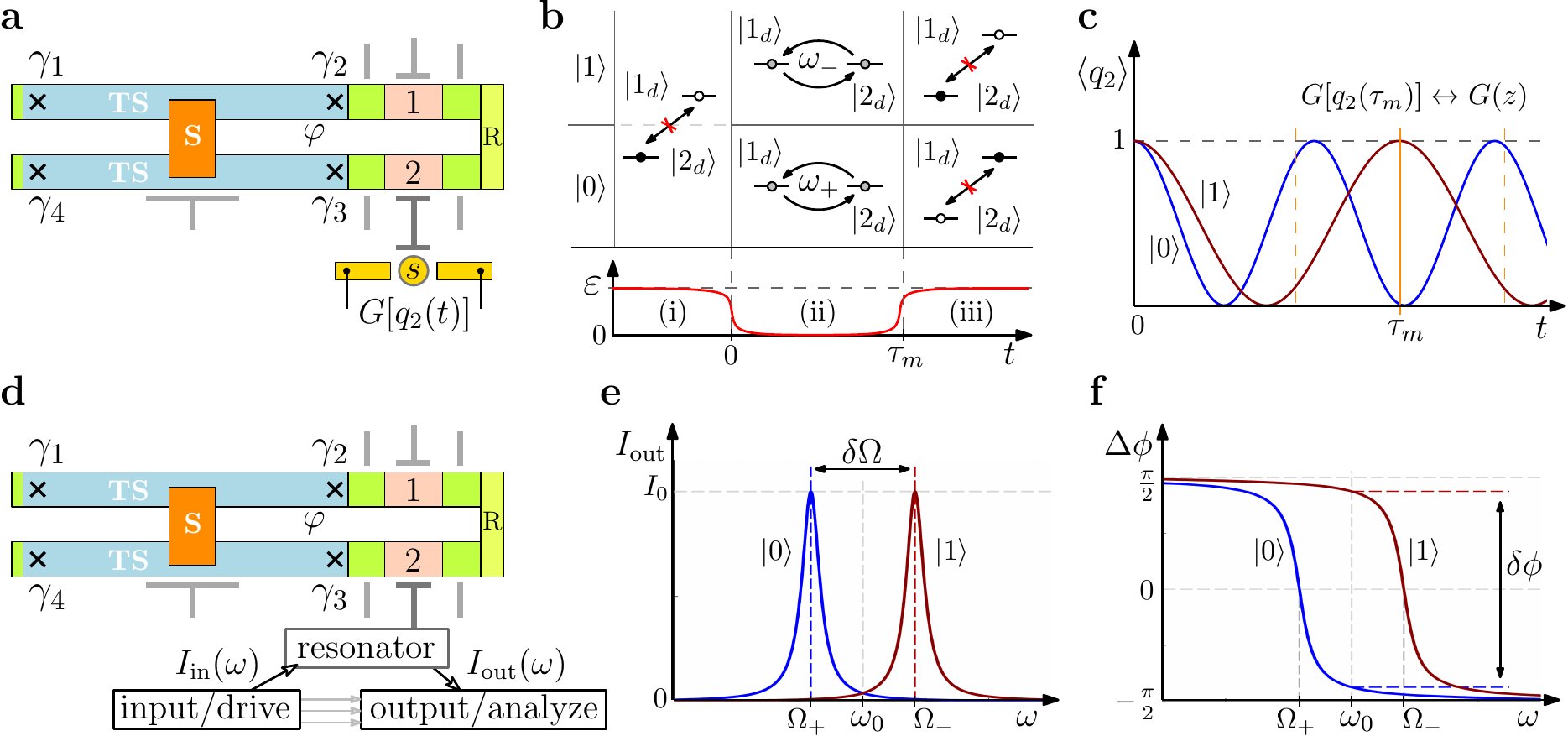}
\caption{\label{fig2}
{
MBQ readout using quantum dots.
{\bf a}, Device for time-domain readout. Two quantum dots 1 and 2 (light red) are formed on top and bottom semiconductor wires (green). Dot levels and tunnel couplings can be adjusted by gates (grey). In addition, dot 2 is capacitively coupled to a charge detector $s$ with conductance $G[q_2]$ depending on dot-2 charge $q_2$.
{\bf b}, Readout protocol. (i) $\varepsilon\gg |t_0\pm t_1|$ at times $t<0$ such that dot 2 is occupied, while dot 1 is empty. (ii) A sudden gate switch at $t=0$ brings the dot levels to resonance, and the electron subsequently undergoes Rabi oscillations between the dots, with frequency $\omega_z$ depending on the qubit state $z=\pm 1$. (iii) After wait time $\tau_m$, one diabatically switches $\varepsilon$ back to a large off-resonant value and measures $q_2$.
{\bf c}, With $\langle q_2(t,z)\rangle \simeq \cos^2(\omega_z t)$, by careful choice of $\tau_m$, e.g., to a maximum of the slower Rabi oscillations, the sensor conductance $G[q_2(\tau_m)]$ will be perfectly correlated with the qubit state $z$ which is thereby read out. For other values of $\tau_m$ (dashed), readout is not ideal.
{\bf d}, Device for frequency-domain readout using charge reflectometry, where a resonator replaces the charge sensor in (a).
{The input signal $I_\mathrm{in}(\omega)$ is either reflected back, or transmitted as output $I_\mathrm{out}(\omega)$ and subsequently analyzed.}
{\bf e}, {When irradiating the resonator with low-bandwidth input $\sim I_0$, a peak in the transmitted photon spectrum $I_\mathrm{out}(\omega)$ for $\omega = \Omega_\pm$ indicates the respective MBQ state $|0\rangle$ or $|1\rangle$.}
{\bf f}, Alternatively, one could measure the $z$-dependent phase shift {$\Delta \phi(\omega)$} of the {transmitted signal, where the contrast $\delta\phi$ is maximized for resonant drive $\omega = \omega_0$.}
}
}
\end{figure*}

The MBQ has four MBSs with corresponding Majorana operators $\gamma_j=\gamma_j^\dagger$. Under strong Coulomb blockade, the fermion parity of the island is a good quantum number, $\gamma_1\gamma_2\gamma_3\gamma_4=\pm 1$, and with negligible Majorana overlaps the four Majorana operators correspond to a degenerate spin-$1/2$ degree of freedom \cite{Beri2012,Altland2013,Landau2016,Plugge2016}. Pauli operators for the MBQ can be represented as
\begin{equation}\label{pauli}
  \X=i\gamma_1\gamma_2,\quad\Y=i\gamma_3\gamma_1,\quad\Z=i\gamma_2\gamma_3,
\end{equation}
expressing the fractionalization of spin into spatially separated MBSs. This spin also appears in the topological Kondo effect \cite{Beri2012,Altland2013} and enters the definition of stabilizers in Majorana surface codes \cite{Landau2016,Plugge2016}. Its non-local topological origin suggests excellent qubit properties, where the MBQ state can be addressed by electron tunneling via weak links between MBSs and either normal leads, Fig.~\ref{fig1}(a), or quantum dots, Figs.~\ref{fig2}(a,d).   Coulomb blockade permits only cotunneling processes, where an electron enters the box through tunneling link $i$ and exits via link $j$. The effective tunneling amplitude then contains a Majorana bilinear $i\gamma_{i}\gamma_j$ \cite{Beri2012,Altland2013,Landau2016,Plugge2016} amounting to one of the Pauli operators in equation \eqref{pauli}.

The simplest MBQ experiment involves interferometric conductance measurements, which provide a natural way to qubit readout and/or initialization in the Pauli eigenbasis \cite{Landau2016,Plugge2016}. The setup is shown in Fig.~\ref{fig1}(a) and we model it by the Hamiltonian
\begin{equation}\label{Ha}
H_a =  H_\mathrm{leads}+\left[\left(t_0+t_1 \Z\right)  d^\dagger_{1} d^{}_{2} + \hc\right],
\end{equation}
where $H_\mathrm{leads}$ describes the two uncoupled leads with density of states $\nu_{1,2}$ and electron operators $d_{1,2}$ near the respective tunnel contact. The cotunneling amplitude via the MBQ is $t_1\Z$, and $t_0$ refers to tunneling through the reference arm in Fig.~\ref{fig1}(a).

In \ref{CondreadoutApp}, we discuss measurement-induced decoherence \cite{Bonderson2007} in this setup, showing that on time scales $t>1/V$, where $V$ is the bias voltage, conductance measurements are projective. There are two possible conductance outcomes, cf.~Fig.~\ref{fig1}(b),
\begin{equation}\label{conductance1}
G_{z}=\frac{e^2}{h}\nu_1\nu_2\left|t_0+ t_1 z\right|^2,
\end{equation}
and the interference term enables readout of the $\Z$-eigenvalue $z=\pm 1$. After the measurement, the MBQ is prepared in the eigenstate determined by the conductance outcome. Since phase coherence in the reference arm requires small $V$, and cotunneling conductances are small, we expect the current-based readout schemes to be limited by the time needed for data accumulation.

We next discuss quantum-dot-based readouts relying on well-known techniques with anticipated faster measurement times compared to the conductance readout. In the setup of Fig.~\ref{fig2}(a), electrons can tunnel back and forth between the two dots either through the box (amplitude $t_1\Z$) or via the reference arm ($t_0$). The corresponding Rabi oscillation period thus depends on the qubit state and its measurement allows for MBQ readout. Similar ideas have been implemented in spin qubit systems \cite{Petta2005} and proposed for single Majorana wires \cite{Hoving2016}. Assuming that the dots have well-resolved single-particle levels, we include only one level per dot (spin-degenerate or spin-polarized). For a single electron occupying the two dots, corresponding to the basis $\{ |1_d\rangle , |2_d\rangle\}$ with detuning energy $\pm \varepsilon$, the system is described by
\begin{equation}\label{Hchargereadout}
  H_b =H_s + \left(\begin{array}{cc} \varepsilon & t_0+t_1\Z \\
  t^*_0+t^*_1\Z \, & \, -\varepsilon + \lambda \hat{Q}_s
  \end{array}\right),
\end{equation}
where $H_s$ models a charge sensor with weak capacitive coupling $\lambda$ between its charge, described by the operator $\hat{Q}_s$, and the dot-2 charge operator $\hat{q}_2=|2_d\rangle\langle 2_d|$.

The readout protocol in Fig.~\ref{fig2}(b) is controlled by $\varepsilon(t)$, i.e., through dot gate voltages. Starting in dot state $|2_d\rangle$ at $t=0$, the electron undergoes Rabi oscillations between the dots (see \ref{SMB}).
Since the Rabi frequency $\omega_{z}=\sqrt{\varepsilon^2+ |t_0+z t_1|^2}$ depends on the qubit state $z=\pm 1$, with carefully timed charge measurements, projective MBQ readout then becomes possible, Fig.~\ref{fig2}(c).
For optimal contrast {(large $\delta\omega = |\omega_+-\omega_-|$),} one can tune $t_0$ and/or the phase $\varphi$ in Fig.~\ref{fig2}(a). The visibility of the Rabi-oscillation readout will of course be reduced by the rather short dephasing time known from similar charge qubits \cite{Nakamura2002}.

Alternatively, we may switch to frequency domain and employ charge reflectometry readout {\cite{Petersson2010,Colless2013,Frey2012,Liu2014}}, for which charge dephasing can be compensated by longer integration times. This setup is illustrated in Fig.~\ref{fig2}(d), where a resonator is capacitively coupled to $q_2$, replacing the sensor dot in Fig.~\ref{fig2}(a). The Hamiltonian $H_c$ is as in equation~(\ref{Hchargereadout}) but $H_s$ now describes the resonator circuit with frequency $\omega_0$, and $\hat{Q}_s = a+a^\dagger$ denotes the coupling to resonator photons $a$. Since the resonator-dot coupling $\lambda$ is weak, we transform from dot basis $\{ |1_d\rangle, |2_d\rangle\}$ to the Rabi basis with Pauli matrices $\tau_{x,y,z}$. In the strong-coupling regime with $\omega_0$ near-resonant with the Rabi frequencies $\omega_{z=\pm}$, the rotating-wave approximation then gives the effective Hamiltonian (\ref{SMB3}, Refs.~\cite{Frey2012,Liu2014})
\begin{equation}\label{HcRWA}
H_c^\mathrm{RWA} = \omega_z\tau_z + \omega_0 a^\dagger a + g_z(a\tau_++a^\dagger\tau_-)~,
\end{equation}
with coupling $g_z = -\lambda|t_0+zt_1|/2\omega_z$. Now adding a drive $H_\mathrm{dr} = i\sqrt{\kappa_\mathrm{in}}[E(t)a^\dagger-E^\ast(t)a]$ on the resonator input port, with single-tone signal $E(t) = \sqrt{I_0} e^{-i\omega t}$ and photon decay rate $\kappa_\mathrm{in}$ into the drive line, the transmitted signal at the output port ($\kappa_\mathrm{out}$) follows from the transmission amplitude $A_\omega = \sqrt{\kappa_\mathrm{out}}\langle a\rangle_{\omega} /\sqrt{I_0}$. Using a master equation approach, we find (\ref{SMB3} and Refs.~\cite{Frey2012,Liu2014})
\begin{eqnarray}\label{transmission}
&& A_\omega = \frac{-i\sqrt{\kappa_\mathrm{in}\kappa_\mathrm{out}}}{ -\frac{i}{2}(\kappa_\mathrm{in}+\kappa_\mathrm{out})+(\omega_0-\omega)+\chi_z}~,\\\nonumber
&& \chi_z = g_z^2/[i\Gamma_\mathrm{tot}-(2\omega_z-\omega)]~.
\end{eqnarray}
The broadening $\Gamma_\mathrm{tot}$ stems from decay and dephasing of the double dot. MBQ readout is now possible either by observing a peak in the amplitude of the transmitted photon spectrum ($I_\mathrm{out}(\omega) = |A_\omega|^2I_0$) at frequency $\omega = \Omega_z$ determined by minimizing $|(\omega_0 - \omega) + \mathrm{Re}(\chi_z)|$ in equation \eqref{transmission}, Fig.~\ref{fig2}(e), or by measuring the {$z$}-dependent phase shift of the transmitted signal ($\Delta\phi(\omega) = -\arg(A_\omega)$), Fig.~\ref{fig2}(f).

As a variant of the quantum-dot-based readout proposed here, we mention the possibility of using the regime where the tunneling through the reference arm ($t_0$) is much stronger than the (co-)tunneling through the MBQ ($t_1$). In this limit, the two dots are effectively hybridized into a single dot tunnel coupled to two Majorana operators, say $\gamma_2$ and $\gamma_3$. The energy shift of the quantum dot depends on $\hat{z}=i\gamma_2\gamma_3$ which therefore can be read out by a measurement of the dot charge \cite{Flensberg2011} or the quantum capacitance \cite{Karzig2016}.

At this point, it is worth stressing that all the above readout schemes are topologically protected in the sense that imperfections that may reduce the readout fidelity (which can be compensated for by longer integration times) do not change the the projection caused by the measurement. This is because the measure operator is uniquely defined by the dots or leads being addressed. The robustness of the projection is a consequence of the non-local and fractionalized nature of the MBQ quantum spin.

\begin{figure}
\hspace{3cm}{\includegraphics[width=\columnwidth]{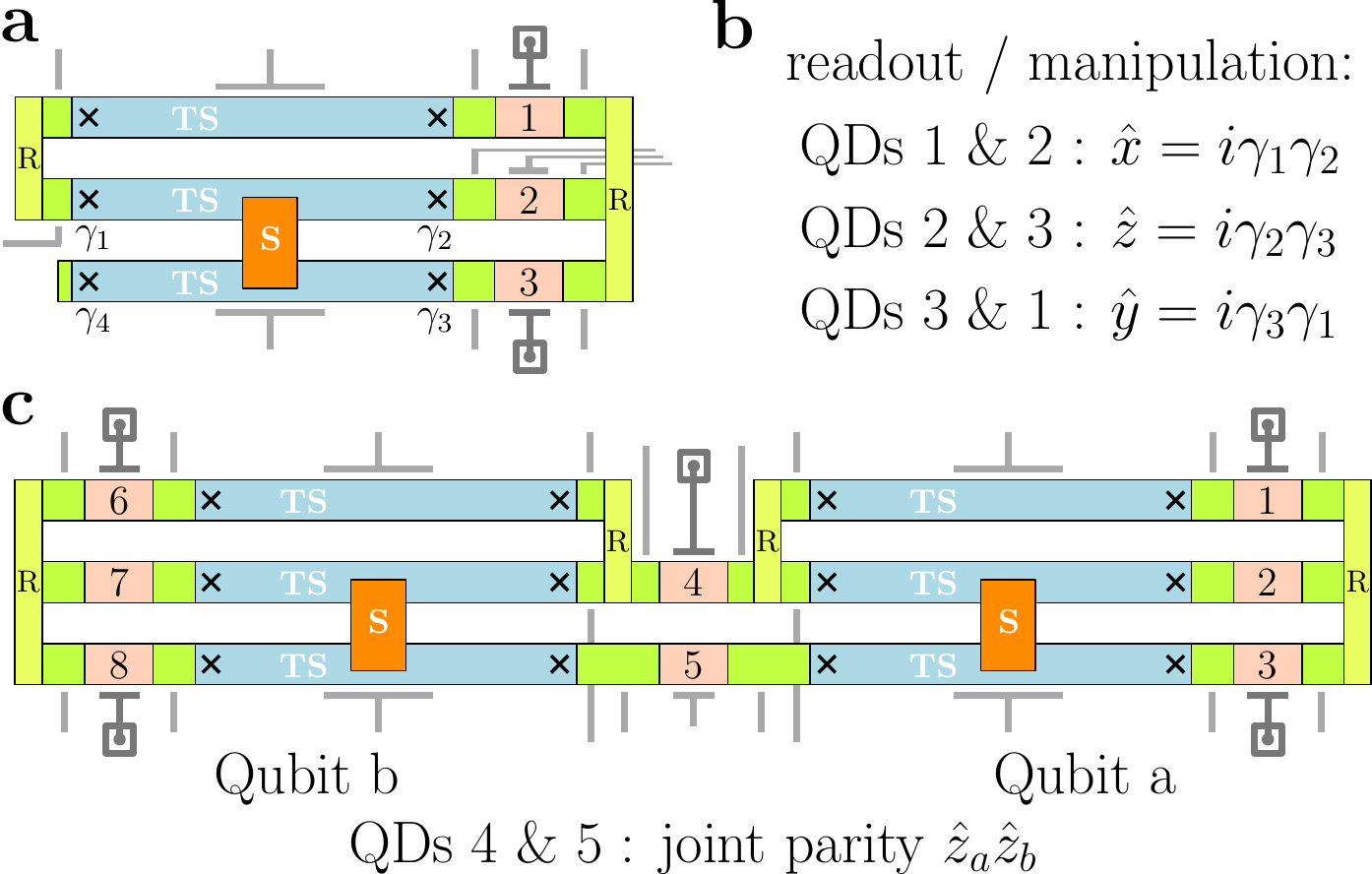}}
\caption{\label{fig3}
Single- and two-qubit devices. {\bf a}, MBQ with three quantum dots and an interference link for readout of all Pauli operators and full one-qubit control. Dark squares indicate either a charge sensor or a resonator system, see~Fig.~\ref{fig2}. {\bf b}, Possible combinations of active dot pairs addressing particular Pauli operators, cf.~equation~\eqref{pauli}. {\bf c}, Device with two MBQs $a$ and $b$ connected by dots 4 and 5, allowing for readout of their joint parity via the MBQ product operator $\Z_a\Z_b$. The other dots serve to read and manipulate qubits individually.}
\end{figure}

So far we discussed readout and preparation of $\Z$-eigenstates. Using the three-dot device with an interference link in Fig.~\ref{fig3}(a), the $\Z$-measurement is readily generalized to readout of all three Pauli operators ($\X, \Y, \Z$). Here, a phase-coherent reference arm connecting far ends of the box is needed, e.g., between $\gamma_1$ and $\gamma_2$.  For this purpose, a floating TS wire (top) acts as a single fermion level stretched out over the entire wire length \cite{Semenoff2006,Fu2010}. Thereby, readout and manipulations along the far side of the MBQ become possible. Fig.~\ref{fig3}(b) lists the corresponding dot pairs to access all Pauli operators. This simple geometry allows for nontrivial test experiments, e.g., to first prepare an eigenstate in one basis, and then measure a different Pauli operator.

Similar protocols allow to manipulate arbitrary MBQ states $|\psi\rangle$. For instance, consider an electron transfer from dot $2\to 3$ in Fig.~\ref{fig3}(a), implemented by ramping the detuning parameter $\varepsilon$. With interference links turned off ($t_0= 0$), the tunneling amplitude is $t_1\Z$, see~equation~\eqref{Hchargereadout}. The protocol begins with an electron on dot 2, $|\Psi(0)\rangle=|\psi\rangle \otimes | 2_d\rangle$.  Assuming that
a later measurement detects an electron on dot 3, the final state is
\begin{equation}\label{etransfer}{
|\Psi_f\rangle= |3_d\rangle\langle 3_d| \left( \mathrm{T}_t\left[e^{-i\int_0^t H dt'}\right]|\Psi(0)\rangle \right)
= (\Z|\psi\rangle) \otimes |3_d\rangle.}
\end{equation}
In effect, the Pauli-$\Z$ operator has thus been applied, $|\psi\rangle\to \Z|\psi\rangle$. Equation \eqref{etransfer} holds because all odd-in-$t_1$ terms are proportional to $\Z$ and because the final measurement has confirmed the transfer $2\to 3$. This protocol works beyond the adiabatic regime \cite{Landau2016,Plugge2016} and allows for fast high-fidelity operations. Moreover, after a failed transfer attempt, $|\Psi_f'\rangle= |2_d\rangle\langle 2_d|(|\Psi(t)\rangle) = |\Psi(0)\rangle$, one can simply retry. Likewise, other Pauli operators are accessible, see Fig.~\ref{fig3}(b). Such manipulations are protected and, without any fine tuning, uniquely determined by the initial and final dot occupations.

\begin{figure}
\hspace{3cm}
\includegraphics[width=.95\columnwidth]{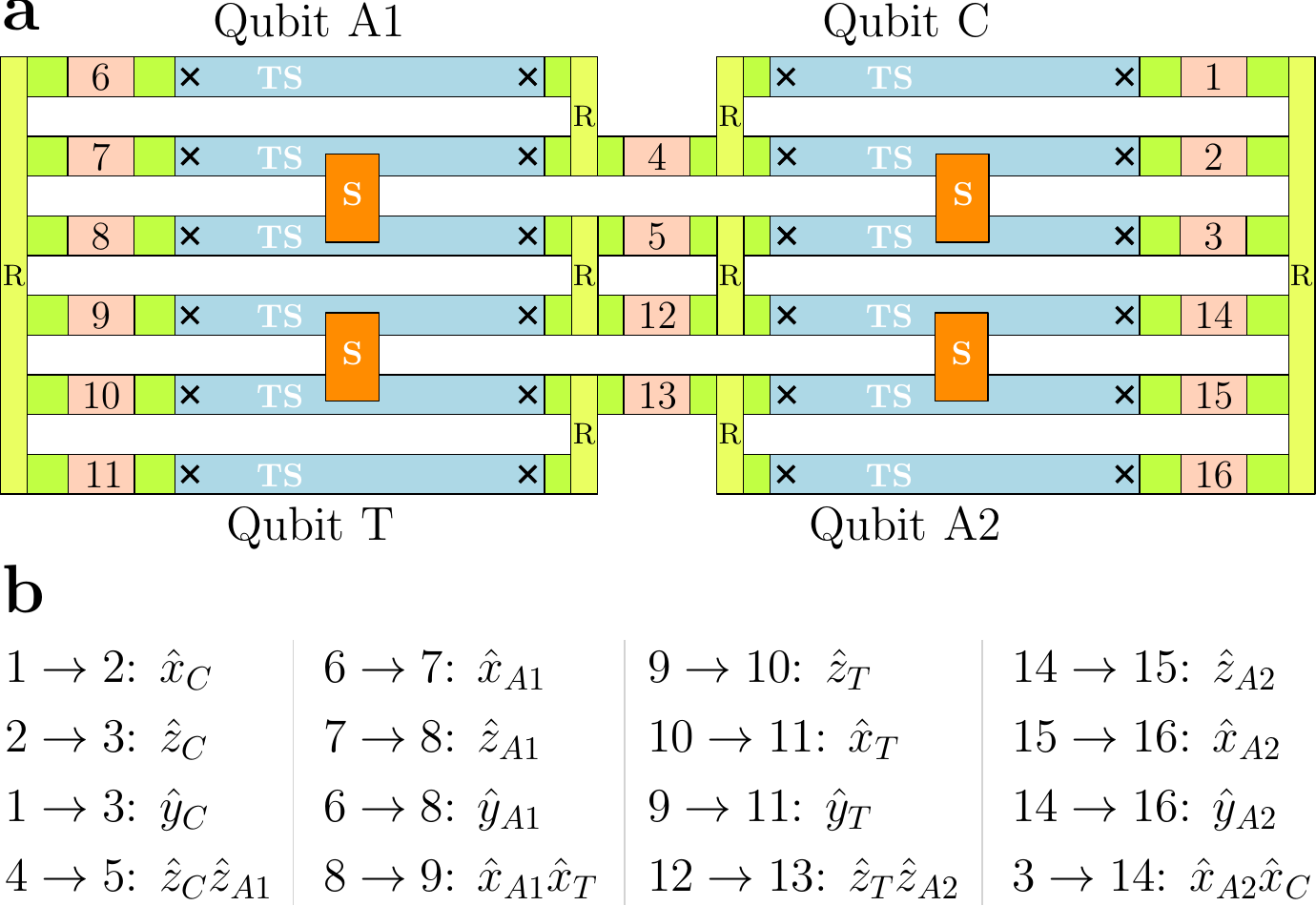}
\caption{\label{fig4}
Four-qubit device. {\bf a}, Similar to Fig.~\ref{fig3}(c) but with four MBQs allowing for implementation of measurement-based topologically protected Clifford group operations, see Fig.~\ref{fig5}. Two data qubits, denoted target (T) and control (C) for a CNOT implementation in Fig. \ref{fig5}(a), are coupled and manipulated by two ancilla qubits A1 and A2 (in the CNOT, A1 serves as active ancilla). The choice of data and ancilla qubits is arbitrary and can be freely interchanged. Using the two ancilla qubits, one can implement $\pi/2$-rotations (i.e., $\hat S$-gates) around both qubit axes $\Z$ and $\X$ of both data qubits C and T. {\bf b}, Extension of the protocols in Figs.~\ref{fig3}(b,c). With the indicated dot pairs, any single-qubit operator and the product operators of adjacent qubits can be addressed.}
\end{figure}

\begin{figure}
\hspace{3cm}{\includegraphics[width=.95\columnwidth]{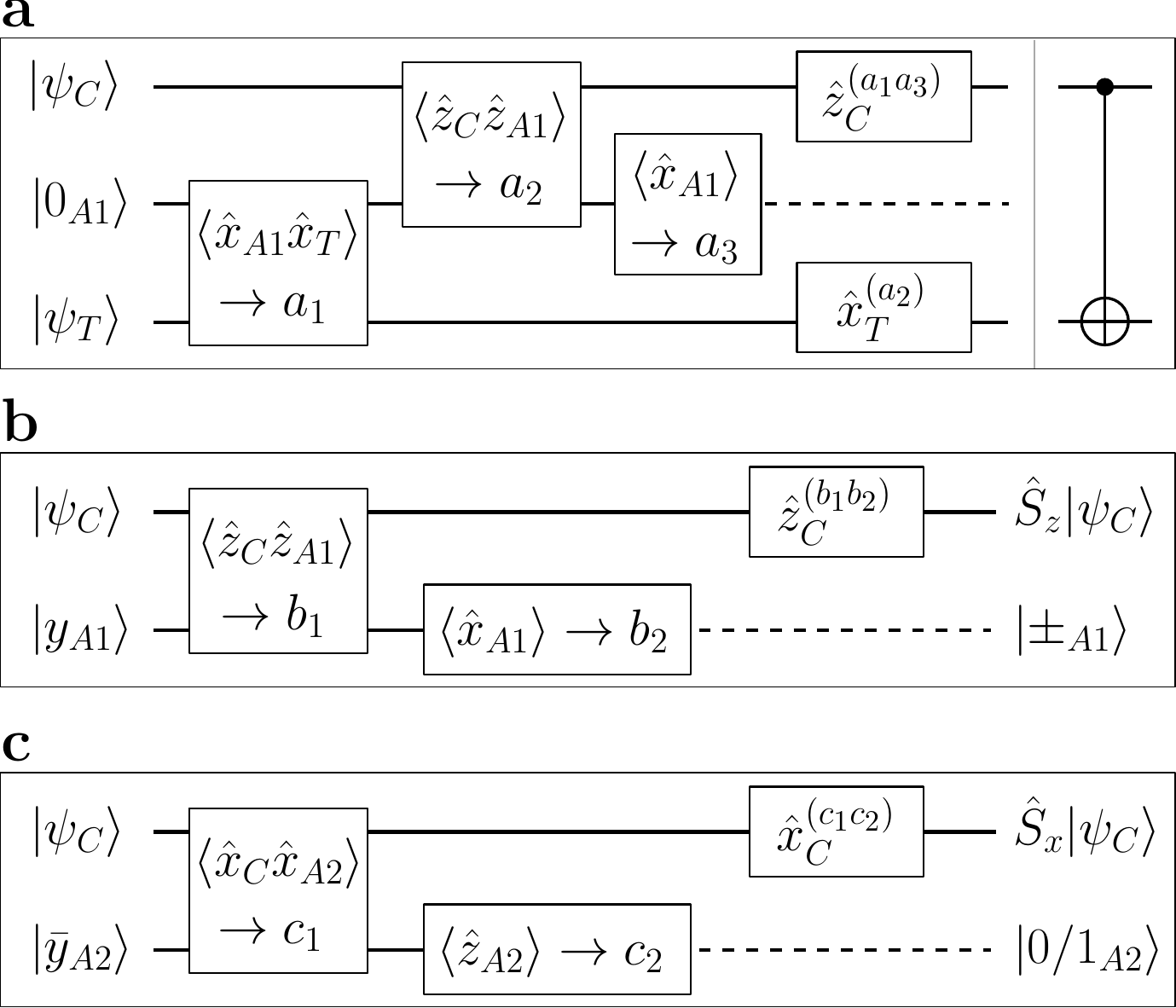}}
\caption{\label{fig5}
Logic circuits for implementation of the Clifford gates.
{\bf a}, Protocol realizing the CNOT gate \cite{Preskill}. After initializing qubit A1 in $|0_{A1}\rangle$, one measures joint parities $\langle\X_{A1} \X_T\rangle$ and $\langle\Z_C \Z_{A1}\rangle$ with respective results $a_1=\pm 1$ and $a_2=\pm 1$. Finally, A1 is read out, $\langle\X_{A1}\rangle = a_3=\pm 1$, followed by controlled Pauli flips on C and T. These flips are conditioned on the intermediate measurement outcomes $a_{1,2,3}$, where $\Z_C$ and $\X_T$ is not applied (is applied) for $a_1a_3 = +1$ $(-1)$ and $a_2 = +1$ $(-1)$, respectively. With these recovery operations, the protocol is guaranteed to give a CNOT gate, cf. inset on the right.
{\bf b}, Protocol implementing $\pi/2$-rotations around the $\Z$-axis on qubit C. After preparing a $y_{A1} = +1$ eigenstate $|y_{A1}\rangle$ on A1, a measurement $\langle\Z_C\Z_{A1}\rangle = b_1$ entangles both qubits. Subsequently, readout of $\langle \Z_{A1}\rangle = b_2$ collapses the state on A1. With recovery operation $\Z_C$ not applied (applied) for $b_1b_2 = +1$ $(-1)$, the protocol implements the desired gate, $\hat S_z = \mathrm{diag}(1,i) \simeq e^{-i\pi\Z/4}$.
{\bf c}, As (b), but for $\pi/2$-rotations around the $\X$-axis.
After preparing a $y_{A2} = -1$ eigenstate $|\bar{y}_{A2}\rangle$ on A2, one measures $\langle\X_C\X_{A2}\rangle = c_1$ and subsequently reads out A2, $\langle \Z_{A2}\rangle = c_2$. With recovery $\X_C$ not applied (applied) for $c_1c_2 = +1$ $(-1)$, the protocol implements the gate $\hat S_x = e^{-i\pi\X/4}$ on qubit C.
Exchanging C $\leftrightarrow$ T and A1 $\leftrightarrow$ A2 in above protocols generates a flipped CNOT (control $\leftrightarrow$ target) and $\pi/2$-rotations on qubit T.
}
\end{figure}

Arbitrary single-qubit rotations are generally not protected to such a degree. Nevertheless, with interference links and fine tuning of tunneling phases, e.g., via the flux $\varphi$ in Fig.~\ref{fig2}, semi-protected operations are possible \cite{Flensberg2011}. Consider dots 2 and 3 in Fig.~\ref{fig3}(a), modelled by equation~\eqref{Hchargereadout} with $\hat t_z= t_0+t_1\Z$. For Re$(t_0^\ast t_1) = 0$,  since $|t_z|$ is independent of $z=\pm 1$,
the MBQ degeneracy remains intact and no dynamical phase is picked up during the electron transfer. One thereby obtains a phase gate, $|\psi\rangle\to e^{i\theta\Z}|\psi\rangle$, with $\theta=\tan^{-1}[$Im$(t_1/t_0)]$.
When combined with other dot pairs, arbitrary rotations are possible. Without phase tuning, protection may be achieved by a four-step pumping protocol \cite{Plugge2016}. In both cases, projective dot charge measurements can eliminate diabatic errors.

A quantum computer is universal if one has full one-qubit control and a two-qubit entangling gate \cite{Nielsen2000}. The above single-qubit measurements can easily be extended to the joint readout of two qubits, which makes the design ideal for measurement-based {gate} protocols \cite{Preskill,Nielsen2000,Bonderson2008}. We first consider an entangling measurement for the two-qubit device in Fig.~\ref{fig3}(c). (See \ref{JointParityApp} for additional details.) With interference links turned off, an electron transfer process from dot $4 \to 5$ has the tunneling amplitude $\hat{t}_{ab} = t_a \Z_a + t_b \Z_b$, where $t_{a/b}$ represents cotunneling via MBQ a$/$b. The corresponding Rabi frequency, $\omega_{z_a z_b=\pm} = \sqrt{\varepsilon^2 + |t_{ab}|^2}$, depends solely on the joint parity $\langle \Z_a\Z_b\rangle=\pm 1$, which can thus be read out as in Fig.~\ref{fig2}. After preparation of $\X_{a,b}$-eigenstates, this operation yields an entangled two-qubit state, and subsequent readout of $\Z_{a,b}$ and/or $\X_{a,b}$ can detect Bell-type correlations \cite{Nielsen2000}.

Finally, the device in Fig.~\ref{fig4} allows for measurement-based topologically protected implementation of the Clifford gates $\lbrace \hat C, \hat S_z, \hat H\rbrace$, see Fig.~\ref{fig5} for the relevant logic circuits.
First, the controlled-NOT (CNOT) gate acts as $|\psi_C\rangle\otimes |\psi_T\rangle\to \hat C(|\psi_C\rangle\otimes|\psi_T\rangle)$, where the operator $\hat C= \frac12(\hat 1+\Z)_C\otimes \hat 1_T+\frac12(\hat 1-\Z)_C\otimes \X_T$ flips the target (T) qubit if and only if the control (C) qubit  is in the $|1\rangle$-state, cf. Fig.~\ref{fig5}(a). Next, single-qubit gates $\hat S_z = \mathrm{diag}(1,i)$ and $\hat S_x =  e^{-i\pi\X/4}$ in Fig.~\ref{fig5}(b,c) together allow for $\pi/2$-rotations around any qubit axis. The Hadamard gate $\hat H = (\X +\Z)/\sqrt2$, effectively exchanging $\X$- and $\Z$-eigenstates, then follows by combined rotations $\hat H = \hat S_z \hat S_x \hat S_z$.

In conclusion, we have described readout, initialization, manipulation, and entangling operations for Majorana box qubits, all of which can be tested in current state-of-the-art experiments. Using these tools, we devised a two-qubit universal quantum computer with protected Clifford gates.  Naturally, the performance of the setup will depend on the fidelity of the readout operation. We have argued that one expects high fidelities because of the topological nature of the qubits, but this of course needs to be confirmed by experimental implementation. Successful demonstration of our proposed devices could pave the way towards fault-tolerant scalable quantum computation, e.g., using surface code architectures and/or hybrid strategies. Finally, we note that interesting generalizations with six or more MBSs on the box could implement measurement-induced braiding operations on a single box \cite{Bonderson2008,Karzig2016}.

\acknowledgements

We thank J. Alicea, A. Altland, P. Bonderson, M. Freedman, L.P. Kouwenhoven, F. Kuemmeth, L.A. Landau, R.M. Lutchyn, C.M. Marcus, C. Nayak, K.D. Petersson, D. Reilly, M.S. Rudner and E. Sela for useful discussions. We acknowledge funding by the Deutsche Forschungsgemeinschaft (Bonn) within the network CRC TR 183 (project C01) and by the Danish National Research Foundation.

\appendix

\section{Conductance-interferometric readout}
\label{CondreadoutApp}

Here we briefly discuss under which conditions the interferometric conductance readout of the MBQ in Fig.~\ref{fig1}(a) will implement a projective measurement. Assuming that the amplitudes $t_{0,1}$ in equation~\eqref{Ha} are switched on at initial time $t=0$ and that the initial state of the MBQ is $|\psi\rangle=\alpha|0\rangle+\beta|1\rangle$,  we obtain for the time-dependent reduced density matrix of the qubit
\begin{equation}\label{rhot}
\rho_\mathrm{MBQ}^{{}}(t) =\left(  \begin{array}{cc}
|\alpha|^2 & \alpha\beta^*e^{-F(t)} \\
\alpha^*\beta e^{-F^*(t)} & |\beta|^2 \\
\end{array}\right).
\end{equation}
The decay of off-diagonal elements is encoded in the real part of the decoherence function
\begin{equation}\label{Ft}
  {\rm Re} F(t)=  4\nu_1\nu_2|t_1|^2 \times \left\{   \begin{array}{cc}
  2\ln (Dt) + \frac12(Vt)^2 ,& \frac{1}{D}\ll t< \frac{1}{V}, \\
   \pi Vt , &  t> \frac{1}{V},
\end{array}\right.
\end{equation}
where $V$ is the applied bias voltage, and the leads have density of states $\nu_{1,2}$ and bandwidth $D$. Once the off-diagonal elements have died out, the conductance measurement can therefore be considered projective. The MBQ will then either be in state $|0\rangle$, with probability $|\alpha|^2$, or in state $|1\rangle$, with probability $|\beta|^2=1-|\alpha|^2$. From the decoherence function $F(t)$ we can thus extract the minimal measurement time needed for a readout based on conductance interferometry, cf. equations (2) and (3) of the main text.

We now turn to the derivation of the above result. We first write the operators $d_{\ell = 1,2}$ at the contact points to the MBQ in terms of conventional lead fermion modes, $d_\ell = \sum_k c_{\ell,k}$. The uncoupled leads Hamiltonian in equation (2) then follows as
$H_{\mathrm{leads}} = \sum_{\ell, k} \xi_{\ell,k} c_{\ell,k}^\dagger c_{\ell,k}~,$
where $\xi_{\ell,k}$ is the occupation energy of lead state $(\ell,k)$.
The reduced density matrix can then be calculated by time-evolving states through
$H_a$ in equation (2), with
$\hat U(t) = \mathrm{T}_t \exp\left[-i\int_0^t H_a(t')dt'\right],$
where $\mathrm{T}_t$ is the time-ordering operator,
followed by performing a trace over the lead degrees of freedom. In our case, $H_a(t)$ is time-independent after initial switch-on of tunnel couplings. Prior to the measurement, the MBQ is detached from the leads and therefore we assume an initial density matrix in product form, $\rho(t=0) =  \rho_\mathrm{MBQ}(t=0)\otimes \rho_\mathrm{leads}$, with initial MBQ density matrix $\rho_\mathrm{MBQ}(t=0) = \sum_{i,j}c_{ij}|i\rangle\langle j|$. The leads are in a thermal state, $\rho_\mathrm{leads} \sim e^{-\beta H_\mathrm{leads}}$ with $\beta = 1/T$, where {the applied bias voltage $V$ determines the chemical potential difference.}
With these ingredients, we obtain the reduced density matrix
\begin{eqnarray}\label{rho_t}
\rho_{\mathrm{MBQ}}(t)
&= \mathrm{Tr}_{\mathrm{leads}}\left[\hat{U}(t)\left(\sum_{i,j}c_{ij}\ket{i}\bra{j}\otimes\rho_{\mathrm{leads}}\right) \hat{U}^\dagger(t) \right] \nonumber \\
& = \sum_{i,j} c_{ij} \ket{i}\bra{j} \avg{U_j^\dagger(t) U_i(t)}_{\mathrm{leads}}.
\end{eqnarray}
Here $U_{i}(t)$ follows by substituting the MBQ Pauli operator $\hat z$ in $H_a$ with its eigenvalue, $\hat z |i\rangle = z_i|i\rangle$ and $\hat U(t)|i\rangle = U_i(t)|i\rangle$, and similar for $U_j^\dagger(t)$. This is possible since the MBQ-mediated tunneling Hamiltonian $H_a$ generically contains only a single Pauli operator per MBQ (in this case $\hat z$), such that the MBQ density matrix can be expressed in the corresponding eigenbasis, here $|i/j\rangle = |0\rangle,~|1\rangle$. It is also clear that diagonal elements (with respect to MBQ tunneling in $H_a$) are conserved, $U_i(t)U_i^\dagger(t) = 1$.
Since the off-diagonal density matrix elements decay,  we obtain
$\rho_\mathrm{MBQ}(t\to\infty) =$ diag$(|c_{00}|^2,|c_{11}|^2)$, cf. equations \eqref{rhot} and \eqref{Ft}.

To obtain an expression for the decay of off-diagonal elements, {we take the lead trace and perform a second-order cumulant expansion in equation \eqref{rho_t}. Considering only the real part (responsible for decay) and noting that elements $\langle 0|\rho_\mathrm{MBQ}|1\rangle$ and $\langle 1|\rho_\mathrm{MBQ}|0\rangle$ are related by Hermitian conjugation, we find
\begin{eqnarray}\label{Ftequation}
&\mathrm{Re} F(t) = -\mathrm{Re} \ln \avg{U_1^\dagger(t) U_0(t)}_{\mathrm{leads}}=\\
 &4\nu_1\nu_2|t_1|^2  \sum_{s = \pm} \int \dintc{\omega} n_B(\omega + sV) \frac{\omega + sV}{\omega^2}(1-\cos \omega t),\nonumber
\end{eqnarray}
where $n_B(\omega)$ is the Bose-Einstein distribution function.
We now evaluate this expression at zero temperature, $n_B(\omega) \to -\Theta(-\omega)$, and introduce a lead bandwidth $D$ as frequency cutoff, $|\omega| \leq D$.
The integrals in equation~\eqref{Ftequation}, summing over $s=\pm$, can then be simplified to give
\begin{eqnarray}\label{integral}
g(t)=\int_{V}^{D}\dintc{\omega}\frac{2(1-\cos(\omega t))}{\omega} +
V\int_{-V}^{V}\dintc{\omega}\frac{1-\cos(\omega t)}{\omega^2}.
\end{eqnarray}
The short-time decay (with $t\ll D^{-1},~V^{-1}$) is Gaussian, $g(t) \approx \frac12(V^2+D^2)t^2$,
while for intermediate times, $D^{-1} \ll t < V^{-1}$ we have $g(t) \simeq ~~2\ln(Dt) +\frac12 V^2t^2~$.
The (asymptotic) for \(t > V^{-1}\) the behavior is $g(t) \simeq ~~2\ln(D/V) + \pi Vt -2(1-\cos(Vt))~$,
where in equation \eqref{Ft} we discarded constant/oscillating parts and kept only the term $\sim t$.

The absolute value of the cotunneling amplitude is given by $|t_1| \simeq |\lambda_1\lambda_2|/E_C$, with lead-Majorana tunnel couplings $\lambda_\ell$ and island charging energy $E_C$ \cite{Beri2012}. The timescale for decay of the off-diagonal elements (i.e., dephasing) of the MBQ state then follows from $\tau_\varphi^{-1}= (\Gamma_1\Gamma_2/E_C^2) V$, containing the lead-island broadenings $\Gamma_{\ell = 1,2} = 2\nu_\ell |\lambda_\ell|^2$ weighted vs the charging energy $E_C$, and
the bias voltage $V$ itself. Since we have to be in the cotunneling regime with $\Gamma_{1,2},V\ll E_C$, for typical device parameters \cite{Albrecht2016},
we find $\tau_\varphi \approx 10~$ns. Conductance readout therefore is most likely
not limited by $\tau_\varphi$ but rather by
data allocation towards sufficient signal-to-noise ratio to resolve the two conductance outcomes, cf.~equation (3) and discussion in main text. Under Coulomb valley
conditions (where our MBQ is operated) the corresponding cotunneling currents are small \cite{Albrecht2016}.\\
Finally, we note that the interference link with tunneling amplitude $t_0$ outside of the MBQ (cf.~equation (2)) does not affect the dephasing time of the MBQ, see equation \eqref{Ftequation}. Here,
tracing out the leads in equation \eqref{rho_t} is the crucial step, where we assume that no knowledge of the leads state is retained, in particular not about the number parity of transferred electrons. For $t_0=0$, when  this number parity is fixed by a measurement, decoherence will not set in because the parity and the number of applications of the corresponding Pauli operator  are perfectly correlated. This fact is exploited by the quantum dot readout and confirmed charge transfer schemes in the main text. With interference links turned off, such protocols
do not involve dephasing of the MBQ state (cf.~equation (7)) but  instead correspond to the application of a Pauli operator to the MBQ state
if an electron has been transferred.
The same statement holds true for the number parity of transferred electrons in the case of normal leads.  If we now include the interference link, $t_0\ne 0$, such correlations between the number parity of transferred electrons and the number of Pauli operator applications will be absent, since the leads are
effectively shorted. In this case, the decoherence is of the more conventional type because the information loss is distributed across a continuum of lead states.  For $t_0=0$, decoherence is only caused by our lack of knowledge about the number parity of transferred electrons.  In any case, the value of $t_0$ does not affect the dephasing time.

\section{Quantum-dot-based readout}
\label{SMB}\label{RabireadoutApp}

Here we discuss details on MBQ readout via (double) quantum dots (QDs), as illustrated in Fig.~\ref{fig2} of the main text. We first translate the bare MBQ plus QDs Hamiltonian into its eigenbasis (\ref{SMB1}), and then analyze the functionality of measurement devices in time-domain (\ref{SMB2}) and frequency-domain readout schemes (\ref{SMB3}).

\subsection{Rabi oscillations for double quantum dot coupled to MBQ}
\label{SMB1}

Since the coupling between dot 2 and the readout device is either intermediately turned off or considered weak, it is convenient to switch to the eigenbasis of $H_b$ in equation~\eqref{Hchargereadout} with $\lambda=0$. We refer to the hybridized double dot as Rabi system. For given Pauli eigenvalue $z=\pm 1$ of the MBQ, the Rabi eigenstates follow from the dot basis $\lbrace |1_d\rangle, |2_d\rangle\rbrace$ as
\begin{eqnarray}\label{RabiBasis}
|\omega_z\rangle &=& \left[t_z|1_d\rangle +(\omega_z-\varepsilon) |2_d\rangle\right]/\sqrt{2\omega_z(\omega_z-\varepsilon)},\\
|\overline{\omega}_z\rangle &=& \left[t_z|1_d\rangle - (\omega_z+\varepsilon)|2_d\rangle\right]/\sqrt{2\omega_z(\omega_z+\varepsilon)},\nonumber
\end{eqnarray}
where $t_z = t_0 +t_1 z$ is total inter-dot tunneling and $\omega_z = \sqrt{\varepsilon^2+|t_z|^2}$ is the Rabi frequency. Since $H_b$ contains a single Pauli operator, the corresponding eigenvalue $z$ is a good quantum number. The same holds true for more participating MBQs where we have a set of good quantum numbers $(z_1,z_2,\ldots)$. We can therefore focus on two Rabi eigenstates $|\omega_z\rangle,~|\overline{\omega}_z\rangle$ throughout, and beyond that refer to the block-diagonal structure of $H_b$ in either quantum dot or Rabi basis. Defining Pauli matrices in the Rabi basis
\begin{equation}\label{RabiPauli}
\tau_z = |\omega_z\rangle \langle\omega_z| - |\overline{\omega}_z\rangle \langle\overline{\omega}_z|,\quad \tau_x = |\omega_z\rangle \langle\overline{\omega}_z| + |\overline{\omega}_z\rangle\langle\omega_z|,
\end{equation}
the bare system Hamiltonian reads $H_0 = \omega_z\tau_z$. Note the hidden $4\times 4$ structure, with two distinct Rabi systems for $z=\pm 1$, cf.~Fig.~\ref{fig2}(b) and equation~\eqref{RabiBasis}.

\subsection{Real-time observation of Rabi oscillations}
\label{SMB2}

For real-time readout of Rabi oscillations on the dots, we first prepare the system at $\varepsilon \gg |t_z|$ locked in the $|2_d\rangle$ state of the dots with charge on dot 2. Next we pulse diabatically to small $\varepsilon$, ideally $\varepsilon \to 0$. The system thus is initialized in the state
\begin{equation}\label{psi0}
|\Psi_0\rangle = \left(\alpha |0\rangle +\beta |1\rangle\right)\otimes |2_d\rangle.
\end{equation}
Switching to the Rabi basis and letting the system evolve up to the wait time $\tau_m$, it will perform Rabi oscillations between the dot states $|2_d\rangle$ and $|1_d\rangle$. The oscillation period depends on the qubit state via the Rabi frequency $\omega_z$, cf. Fig.~2(b) of the main text, leading to an entangled state of MBQ and QDs. The time-evolved state of the QDs is then given by
\begin{equation}\label{evolve}
|\Psi(t)\rangle=
\frac{it_z}{\omega_z}\sin(\omega_z t)|1_d\rangle +
\left(\cos(\omega_z t) - \frac{i\varepsilon}{\omega_z}\sin(\omega_z t)\right)|2_d\rangle~.
\end{equation}
As the next step, we diabatically pulse back to large $\varepsilon$ such that the remaining dynamics follow from effectively decoupled dots. Inserting the wait time $\tau_m$ spent at the measurement point $\varepsilon = 0$, we find the charge measurement outcome probabilities $p_z(q)$
\begin{eqnarray}\label{measure_probability}
p_+(2) &=& |\alpha|^2\left[\cos^2(\omega_+\tau_m)+\frac{\varepsilon^2_m}{\omega_+^2}\sin^2(\omega_+\tau_m)\right],\\ \nonumber
p_+(1) &=& |\alpha|^2\left[\frac{|t_+|^2}{\omega_+^2}\sin^2(\omega_+\tau_m)\right],\\ \nonumber
p_-(2) &=& |\beta|^2\left[\cos^2(\omega_-\tau_m)+\frac{\varepsilon^2_m}{\omega_-^2}\sin^2(\omega_-\tau_m)\right],\\ \nonumber
p_-(1) &=& |\beta|^2\left[\frac{|t_-|^2}{\omega_-^2}\sin^2(\omega_-\tau_m)\right],
\end{eqnarray}
where $q = 1,~2$ indicates whether the charge was found on QD 1 or 2, and $\varepsilon_m$ is the value of the detuning during the time-evolution described by equation \eqref{evolve}. As prescribed by the initial state \eqref{psi0}, we find $p_+(1)+p_+(2) = |\alpha|^2$ and $p_-(1)+p_-(2) = |\beta|^2$. Further, maximum oscillation amplitudes of the probabilities are achieved for $\varepsilon = 0$ at the measurement point, giving
\begin{eqnarray}
\bar p_+(2) = |\alpha|^2\cos^2(\omega_+\tau_m)~,~~\bar p_+(1) = |\alpha|^2\sin^2(\omega_+\tau_m),\nonumber\\
\bar p_-(2) = |\beta|^2\cos^2(\omega_-\tau_m)~,~~\bar p_-(1) = |\beta|^2\sin^2(\omega_-\tau_m),\nonumber
\end{eqnarray}
where the Rabi frequency reduces to $\omega_z(\varepsilon= 0) = |t_z|$. Maximum contrast is achieved for $\tau_m$ where $p_+(2)\ll p_-(2)$ (modulo amplitudes $|\alpha|$ and $|\beta|$). A charge measurement finding $q=2$ thus  identifies the qubit state to be $|1\rangle$ ($z = -1$) with high fidelity. Conversely, finding $q = 1$  with $p_+(1) \gg p_-(1)$ identifies the qubit to be in state  $|0\rangle$ ($z=+1$). As an example, we plot the probabilities $\langle q_2(t)\rangle = \cos^2(\omega_z t)$ (i.e., $\bar p_z(2)$) for initial qubit states $|0\rangle$ and $|1\rangle$ in Fig. 2(c) of the main text.

Let us now consider the action of a dot-charge measurement on the dynamically evolved state $|\Psi(\tau_m)\rangle$ defined through equations \eqref{psi0} and \eqref{evolve}. Assuming the initial state as above ($\varepsilon=0$) and for a measurement finding $q = 2$, the qubit will be in the state
\begin{equation}
|\Psi(\tau_m;2)\rangle = \frac{\alpha\cos(\omega_+\tau_m)|0\rangle + \beta\cos(\omega_-\tau_m)|1\rangle}{\sqrt{\bar p(\tau_m;2)}},
\end{equation}
with total measurement probability $\bar p(\tau_m;2) = \bar p_+(2) + \bar p_-(2)$ as normalization. Conversely, if one measures $q = 1$ the qubit state is
\begin{equation}
|\Psi(\tau_m;1)\rangle = \frac{\alpha e^{i\varphi_+} \sin(\omega_+\tau_m)|0\rangle + \beta e^{i\varphi_-}\sin(\omega_-\tau_m)|1\rangle}{\sqrt{\bar p(\tau_m;1)}},
\end{equation}
with $\bar p(\tau_m;1) = \bar p_+(1) + \bar p_-(1)$. The phases $\varphi_z =  t_z/|t_z| = \arg(t_z)$ stem from the prefactor of the $|1_d\rangle$-term in equation \eqref{evolve}.

Above we saw that the action employed by measuring the dot state $q = 1,~2$ will not fully project the qubit state in the general case. Only when one of the probabilities vanishes is the projection complete, which however requires perfect timing of $\tau_m$. Deviations from this perfect value, together with imperfections due to noise, limits the readout fidelity. Unfortunately, one cannot simply re-measure multiple times to obtain better statistics. Instead, the upper limit to improvements through repetition is set by the visibility of charge oscillations.
In other words, re-measuring with updated coefficients $\alpha' = \alpha\cos(\omega_+\tau_m) / \sqrt{\bar p(\tau_m;2)}$ and $\beta' = \beta\cos(\omega_-\tau_m) / \sqrt{\bar p(\tau_m;2)}$ after initially measuring  $|2_d\rangle$ and repeating will not completely converge the qubit state. For further discussions on measurement details of the time-domain readout method, see e.g. Ref.~\cite{Hoving2016}.
}
\subsection{Charge reflectometry readout}\label{SMB3}

Here we discuss the charge reflectometry readout of MBQ-QD hybrid systems in more detail. As discussed above, readout based on coherent charge oscillations in Fig.~2(b,c) will be limited by charge fluctuations and noise. Driven high-frequency charge oscillations might be more robust in this regard and the technique is well-known in the context of double-dot spin or charge qubits \cite{Frey2012,Liu2014}. As illustrated in Fig.~2(d), a resonator is capacitively coupled to the QD charge to be read out.

We translate the coupling term $\sim \lambda \hat{Q}_s \hat{q}_2$ in equation \eqref{Hchargereadout} into the Rabi basis. Replacing $q_2$, one finds
\begin{equation}\label{Hc}
H_c = H_0 + H_s + \frac12\lambda \hat{Q}_s\left(1-\frac{\varepsilon} {\omega_z}\tau_z-\frac{|t_z|}{\omega_z}\tau_x\right)~.
\end{equation}
Now consider a resonator with bare frequency $\omega_0$, modeled by $H_s = \omega_0 a^\dagger a$, and capacitive coupling to the resonator photons, $\hat{Q}_s = a+a^\dagger$.
When resonator and double dot are near-resonant, $\omega_0 \approx 2\omega_z$ and $|\omega_0-2\omega_z|\ll $ min$(\omega_0,\omega_z,|\omega_0+2\omega_z|)$, the rotating-wave approximation gives $H_c^\mathrm{RWA}$ in equation \eqref{HcRWA} of the main text.

\subsubsection{Strong-coupling regime}\label{strongcoupApp}

First we consider the strong coupling regime (focused on in main text), where the hybridized double dot (Rabi system) is at resonance with the resonator, $\omega_0 \simeq 2\omega_z$. Within the rotating-wave approximation, see equation (5), one can re-express the system in terms of dressed states, see e.g. Ref.~\cite{Blais2004}. Similar calculations as done here for the strong-coupling regime can be found in Refs.~\cite{Petersson2010,Colless2013,Frey2012,Liu2014} along with experiments. In order to allow readout of the energy spectrum of the system we add a drive term $H_\mathrm{dr} = i\sqrt{\kappa_\mathrm{in}}\left[E(t)a^\dagger -E^\ast(t)a\right]$, with field $E(t) = \sqrt{I_0}e^{-i\omega t}$ incident at the resonator. Switching to the rotating frame of the drive, the full Hamiltonian reads
\begin{equation}
H = \frac12 \Delta_z\tau_z + \Delta_0 a^\dagger a + g_z(a\tau_++a^\dagger\tau_-)+ i\sqrt{\kappa_\mathrm{in} I_0}(a^\dagger-a)~,
\end{equation}
where the effective coupling $g_z= -\lambda|t_z|/2\omega_z$ describes emission and absorption of resonator photons. Pauli matrices $\tau_\pm =(\tau_x\pm i\tau_y)/2$ are defined in accordance with equation \eqref{RabiPauli}. The dots and photon energies in the rotating frame are shifted to $\Delta_z = 2\omega_z-\omega$ and $\Delta_0 = \omega_0 -\omega$.\\
The time evolution now follows from a standard master equation approach. In the presence of coupling to baths, the Liouville equation for the density matrix reads
\begin{eqnarray}
\dot{\rho} =& -i\left[H,\rho\right] + \frac{\Gamma_\phi}{2}\mathcal{D}[\tau_z]\rho \\
&+\gamma(n_\mathrm{th}+1)\mathcal{D}[\tau_-]\rho
+\gamma n_\mathrm{th}\mathcal{D}[\tau_+]\rho~\nonumber\\
&+ (\kappa_\mathrm{in}+\kappa_\mathrm{out})\left[(1+N_\mathrm{th})\mathcal{D}[a]\rho+
 N_\mathrm{th}\mathcal{D}[a^\dagger]\rho\right]~,\nonumber
\end{eqnarray}
where $\mathcal{D}[A]\rho = A\rho A^\dagger -\frac12 \lbrace A^\dagger A,\rho\rbrace$ is the Lindblad dissipator. Terms $\sim\kappa_\mathrm{in/out}$ model decay of resonator photons into the input/output line, while $\gamma$ and $\Gamma_\phi$ quantify decay and dephasing of the hybridized double dot. The baths set the temperature $T$ of the system, where $n_\mathrm{th} = 1/(e^{2\omega_z/kT}-1)$ and $N_\mathrm{th}=1/(e^{\omega_0/kT}-1)$. We then find the equations of motion
\begin{eqnarray}
\langle\dot{a}\rangle &=& -i\Delta_0\langle a\rangle -ig_z\langle \tau_-\rangle + \sqrt{\kappa_\mathrm{in}I_0} -\frac12(\kappa_\mathrm{in}+\kappa_\mathrm{out})\langle a\rangle~,\\
\langle\dot{\tau}_-\rangle &=& -i\Delta_z\langle \tau_-\rangle +ig_z\langle a\tau_z\rangle -\Gamma_\mathrm{tot}\langle \tau_-\rangle~,
\nonumber
\end{eqnarray}
with total decay rate
$\Gamma_\mathrm{tot} = \Gamma_\phi + \frac12 \gamma(1+2n_\mathrm{th})$.
For the steady-state solution we take the semiclassical decoupling approximation, $\langle a\tau_z\rangle_\mathrm{ss} \approx \langle a\rangle_\mathrm{ss}\langle \tau_z\rangle_\mathrm{ss}$. Due to external dephasing and decay ($\Gamma_\mathrm{tot}$), once the steady state is reached, the double dot then is assumed to be in a thermal equilibrium with the environment, i.e., $\langle\tau_z\rangle_\mathrm{ss} = -\tanh(\omega_z/kT)$.
We find
\begin{eqnarray}
\langle\tau_-\rangle_\mathrm{ss} &=& -\frac{\chi_z}{g_z}\langle a\rangle_\mathrm{ss}\langle \tau_z\rangle_\mathrm{ss}~,\\
\langle a\rangle_\mathrm{ss} &=& \frac{-i\sqrt{\kappa_\mathrm{in}I_0}}{ -\frac{i}{2}(\kappa_\mathrm{in}+\kappa_\mathrm{out})+\Delta_0-\chi_z\langle \tau_z\rangle_\mathrm{ss}}~,
\nonumber
\end{eqnarray}
where we defined the susceptibility $\chi_z = g_z^2/(i\Gamma_\mathrm{tot}-\Delta_z)$. From here, we directly obtain the transmission amplitude
\begin{equation}\label{transamp}
A_\omega = \frac{\sqrt{\kappa_\mathrm{out}}\langle a\rangle_\mathrm{ss}}{\sqrt{I_0}} = \frac{-i\sqrt{\kappa_\mathrm{in}\kappa_\mathrm{out}}}{ -\frac{i}{2}(\kappa_\mathrm{in}+\kappa_\mathrm{out})+\Delta_0-\chi_z\langle \tau_z\rangle_\mathrm{ss}}.
\end{equation}
In equation \eqref{transmission} of the main text we took for simplicity the zero-temperature limit $\avg{\tau_z}_\mathrm{ss}\to -1$. One sees that the real part of $\chi_z$ describes shifts of the transmission resonances, where a maximum of $|A_\omega|^2$ follows by minimizing $|\Delta_0-\mathrm{Re}(\chi_z)\avg{\tau_z}_\mathrm{ss}|$, while the imaginary part modifies the resonator decay rates.
Finite temperature $T$ diminishes the relative shifts $\sim \chi_z\avg{\tau_z}_\mathrm{ss}$, with $\avg{\tau_z}_\mathrm{ss} \to 0$ for $kT\gg \omega_z$. We plot the transmission $I_\mathrm{out}(\omega) = |A_\omega|^2 I_0$ and the phase shift $\Delta \phi_\omega = -\arg(A_\omega)$ in Figs. \ref{fig2}(e,f) of the main text, where for illustration purposes we take the idealized situation $\Gamma_\mathrm{tot} = 0$ and $T=0$. Other parameters are $\lambda = 2$ (then $|g_z(\varepsilon = 0)| = 1$), $\omega_0 = 10$, $\kappa_{\mathrm{in/out}} = 0.2$ and Rabi frequencies $\omega_+ = 6$, $\omega_- = 4$ in arbitrary units.
In this case $\omega_0$ is centered between the transition frequencies $2\omega_z$ of the double dot. Therefore also the resulting transmission peak shifts ($\Omega_z$) and phase shifts ($\Delta \phi$) are symmetric around $\omega_0$, cf. Figs. \ref{fig2}(e,f).
Other choices of $\omega_{z=\pm}$ relative to $\omega_0$, e.g., by adjusting the detuning $\varepsilon$, may yield better contrast ($\delta\Omega$ and/or $\delta\phi$ in Figs. \ref{fig2}(e,f)) depending on the remaining parameters, cf. equation \eqref{transmission} and the discussion below.
\begin{figure}
\hspace{2.25cm}{\includegraphics[width=.95\columnwidth]{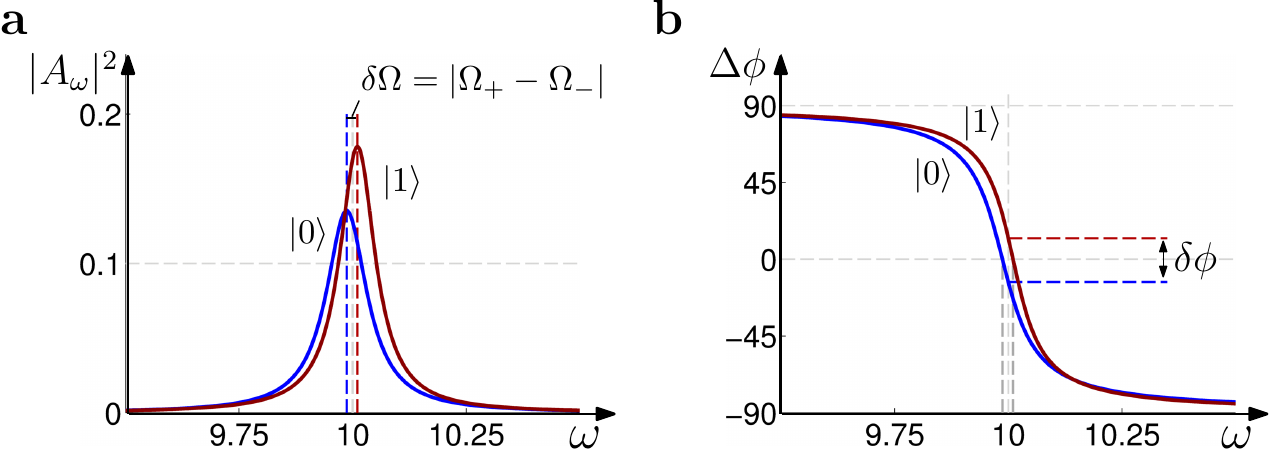}}
	\caption{{\bf a,} Resonator transmission $|A_\omega|^2$ vs. drive frequency $\omega$ [in units $\lambda = 200$ MHz] for zero double-dot detuning, $\varepsilon = 0$ such that $\omega_z = |t_z|$ with remaining parameters given in the text. {Vertical dashed lines indicate the position of principal resonances $\omega = \Omega_z$. Clearly, the signals and splitting $\sim\delta\Omega$ are} diminished by broadening $(\Gamma_\mathrm{tot})$ and finite temperature $(T)$ effects, cf. equation \eqref{transamp}. When going beyond the RWA, we expect to see further (but weaker) dips at larger drive-resonator detuning, where the double-dot splitting crosses multi-photon resonances of the cavity (e.g., a two-photon resonance at $2\omega_0 = 2\omega_z$). {\bf b,} Transmission phase shift $\Delta\phi$ [in degree] vs. drive frequency $\omega$ for same parameters as in (a).}	\label{f1supp}
\end{figure}

Taking experimentally relevant parameters from the cavity - double dot experiments in Refs.~\cite{Frey2012,Liu2014} in combination with Majorana wire experiments~\cite{Albrecht2016} as guideline, we next consider: bare coupling $\lambda = 1$ (then $|g_z(\varepsilon = 0)| = 0.5$), $\omega_0 = 10$, $\kappa = \kappa_\mathrm{in/out}/2 = 0.02$ (i.e., resonator quality factor $Q=\omega_0/\kappa = 500$) and tunnel couplings $|t_+| = |t_0+t_1| = 6$, $|t_-| = |t_0-t_1| = 4$ in units of $200$ MHz. Further, we consider decay and dephasing of the hybridized double dot (Rabi system) as $\gamma = 0.5$ and $\Gamma_\phi = 5$ (cf. \cite{Frey2012,Liu2014}), respectively, and temperature $T = 5$ (corresponding to $1$ GHz $\simeq$ $50$ mK on-chip base temperature). With these parameters the plots in Fig. 2(e,f) change to Fig. \ref{f1supp}(a,b). If the Rabi frequencies are centered around the bare resonator frequency $\omega_0$, one of the frequency shifts is negative ($\Omega_+ < \omega_0$) and the other positive ($\Omega_->\omega_0$). However, they are no longer symmetric around $\omega_0$ which is caused by the broadening and finite-temperature effects. Similarly, the phase shift for the $z=+(-)$ MBQ state at resonant drive $\omega_0 = \omega$ is negative (positive). The point $\Delta\phi = 0$ is crossed when running across the transmission resonance, i.e., for $\omega = \Omega_{+(-)}$, as indicated by the {vertical dashed} lines in the figure. The configuration of energy scales/frequencies with double dot transition energies $2\omega_z$ centered around $\omega_0$ is advantageous for phase shift readout, where the contrast $\delta\phi$ is maximized for a resonant drive $\omega = \omega_0$ matching the bare resonator frequency. Such differential phase shifts of order $\delta\phi \gtrsim 10^\circ$ can be clearly resolved in experiments, cf. \cite{Frey2012}, and readout of the qubit state should be possible with integration times in the sub-$\mu$s regime.

\begin{figure}[t]
\hspace{2.25cm}{
\includegraphics[width=.95\columnwidth]{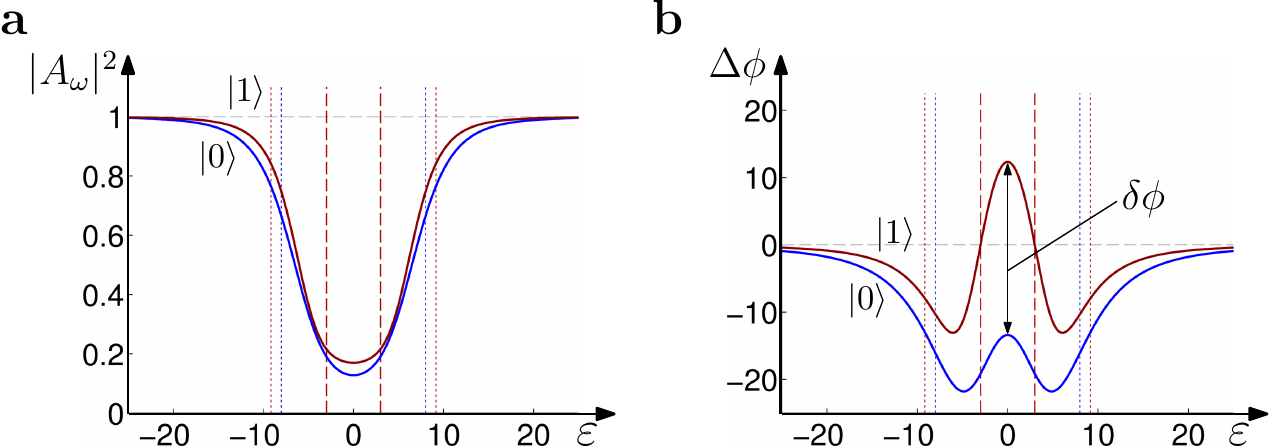}}
	\caption{
	{\bf a,} Resonator transmission $|A_\omega|^2$ vs. double-dot detuning $\varepsilon$ [in units $\lambda = 200$ MHz] for a drive resonant with the bare resonator frequency, $\omega = \omega_0$. Other parameters are as in Fig.~\ref{f1supp}.
	{\bf b,} As (a), but now for the phase shift $\Delta\phi$ [in degrees] of the transmitted signal.
	In both plots the red dashed lines indicate the points where $2\omega_- = \omega_0$, appearing symmetrically around $\varepsilon = 0$ (as is the whole plot) since $\varepsilon$ only enters quadratically in $\omega_z$. Further red(blue) dotted lines denote detuning strengths where multi-photon resonances $2\omega_0 = 2\omega_-$ ($2\omega_0 = 2\omega_+$) are crossed. These processes are not captured within the rotating-wave approximation. We expect such resonances to be much weaker than the principal ones at $\omega_0 \simeq 2\omega_z$, in particular for finite (large) broadening $\Gamma_\mathrm{tot}$ and temperature $T$.
	}
	\label{f2supp}
\end{figure}

Last, we show the resonator response as function of the double-dot detuning $\varepsilon$, see Figs.~\ref{f2supp}(a,b), where the resonator is driven at its bare resonance frequency, $\omega = \omega_0$. As long as the double-dot is far-detuned from the resonator, $|2\omega_z -\omega_0| \gg g_z$, the two systems effectively decouple and the drive signal is transmitted resonantly by the bare resonator, with full transmission $|A_\omega|^2 = 1$, cf. Fig.~\ref{f2supp}(a). As the hybridized double dot energy splitting $2\omega_z$ approaches the bare resonance frequency $\omega_0$, the systems eigenstates are transformed into dressed states, cf. also Refs.~\cite{Frey2012,Liu2014,Blais2004}. Consequently, first due to dispersive and ultimately the strong-coupling shifts of the resonance frequency $\omega_0 \to \Omega_z$, the drive and resonator become off-resonant. The transmission is therefore suppressed at smaller values of $\varepsilon$, cf. Fig.~\ref{f2supp}(a), and a large part of the input signal is back-reflected. In addition, decay and dephasing of the double-dot cause energy loss of the drive, which diminishes the transmitted signal. As can be seen in Fig.~\ref{f2supp}(a), while a measurement of the resonator transmission in principle can give information about the qubit state, the differences for MBQ states $|0\rangle$ and $|1\rangle$ (blue/red curve) are quite small. This is because the readout configuration chosen for Fig.~\ref{f2supp} (and Fig.~\ref{f1supp}) is targeted towards transmission-phase-shift readout, with drive frequency near the resonator frequency, $\omega = \omega_0$.
In Fig.~\ref{f2supp}(b) the phase shift is seen to go to zero when the double dot is far-detuned from the resonator, whereas  when the double dot is close to resonance it pulls the resonator frequency $\omega_0 \to \Omega_z$. Consequently, a finite phase shifts arises with $\Delta\phi < 0$, as long as $2\omega_z$ is larger than the resonator frequency. Since we consider parameters with $2|t_+|> \omega_0$ and $2|t_-|<\omega_0$, for the MBQ state $z=+$ the resonance $2\omega_z = \omega_0$ is never crossed, and the transmission phase shift stays negative for any $\varepsilon$. Conversely, for MBQ state $z = -$ the resonance condition $2\omega_- = \omega_0$ is fulfilled at finite $\varepsilon = \pm \sqrt{\omega_0^2/4-|t_-|^2}$ (red dashed line in Figs.~\ref{f2supp}(a,b)), and the transmission phase shift changes sign. A measurement finding a positive phase shift $\Delta\phi$ at $\varepsilon\to 0$ (for this parameter setting) thus detects the MBQ state $z = -1$, and a negative shift indicates $z=+$. Very similar experiments as suggested here have been successful in detecting such shifts (of similar magnitude as predicted here) when tuning the tunnel coupling of a double dot, cf. Ref.~\cite{Frey2012}. While in the case of Frey \textit{et al.} \cite{Frey2012} the tuning of tunnel couplings was introduced by a gate, here the different tunnel couplings $|t_z|$ (or Rabi frequencies $\omega_z$) encode the MBQ state. It should be noted that the strong coupling regime also has more complex bistability behaviors \cite{Peano2010}, which should be avoided in order not to confuse the readout measurements. Unwanted multi-photon phenomena can of course complicate the readout, but they cannot, as discussed in the main text, alter the operator being projected by the measurement.

\subsubsection{Dispersive regime}\label{SMB32}
Secondly we discuss readout in the dispersive regime, cf. Refs.~\cite{Ohm2015,Yavilberg2015} for similar models and discussion of Majorana systems coupled to microwave resonators. Depending on experimental details, operation of the resonator and readout devices either in the strong-coupling or dispersive regime may be more practical. We again start from $H_c$ in equation \eqref{Hc}, now in the dispersive regime. In this situation, only virtual photon processes are relevant, and expansion to second order in $\lambda/\omega_0$ and $\lambda/(\omega_0\pm2\omega_z)$ yields the effective Hamiltonian (cf.~Refs.~\cite{Ohm2015,Yavilberg2015})
\begin{eqnarray}\label{shift}
&& H_c^{\rm eff}=  \left( \omega_z +\frac{\delta_z}{2} +\frac{\chi_z}{2} \right) \tau_z
+ \left(\omega_0+\chi_z \tau_z \right) a^\dagger a + c_z ,\\\nonumber
&& \delta_z = \frac{\lambda^2\varepsilon}{\omega_0\omega_z},
 \quad \chi_{z} = \frac{\lambda^2|t_0+t_1 z|^2}{\omega_z(\omega_0^2-4\omega_z^2)},\quad c_z=
 -\frac{\varepsilon\delta_z+\omega_0\chi_z}{4\omega_z}.
\end{eqnarray}
The dispersive shift $\chi_z$ becomes large for $\omega_0\to2\omega_z$, where the perturbative expansion breaks down and the photon field resonantly drives Rabi oscillations between the dots (i.e., we are back to the strong-coupling case). From the Hamiltonian \eqref{shift} we can directly read off the shifted resonance frequencies $\Omega_z = \omega_0 +\chi_z\langle\tau_z\rangle_\mathrm{ss}$ when taking the steady-state expectation value $\tau_z \to \langle\tau_z\rangle_\mathrm{ss}$. Note that in this simplified approach neither resonance broadening due to decay and dephasing of the Rabi system ($\Gamma_\mathrm{tot}$) nor decay of the resonator photons ($\kappa_\mathrm{in/out}$) are considered.
Both effects (along with the drive) can be included as for the strong-coupling case following the same steps as above.

Finally, the precise form of the strong-coupling or dispersive shifts $\sim \Omega_z$ depends on the type of coupling between system and resonator. However, we expect that there is no sizable {direct} coupling $\sim |1_d\rangle\langle 2_d| ,~|2_d\rangle\langle 1_d|$ between the QDs apart from tunneling via the interference link or the MBQ. Therefore any resonator or gate couples only to the QD charge operators $\hat{q}_1$ and $\hat{q}_2$, and the resulting shifts appear to contain only information about the tunneling amplitudes $|t_z|$. This property is highly desirable in particular for joint-parity readouts used to entangle adjacent qubits.

\section{Joint-parity measurements}\label{JointParityApp}

In the main text we have discussed single-qubit and joint-parity measurements using quantum dots. For example, for two-qubit readout with tunneling $t_{ab} = t_a z_a + t_b z_b$, see Fig.~3(c),
the readout signal depends through the Rabi frequency $\omega_{z_az_b} = \sqrt{\varepsilon^2+|t_{ab}|^2}$ only on the joint parity $z_az_b = \pm 1$. It is, however, not obvious that the state of the two-qubit system is only projected onto a subspace with $z_az_b=\pm1$, and not further affected within this subspace. In this Appendix, we show that measurement-induced dephasing or accidental qubit rotations are completely avoided, a property inherited from the geometric protection of {MBQ} spins. Any single-qubit and joint-parity measurements performed by these methods thus are expected to offer exceptionally high readout fidelities and low probability of readout-induced errors. We also note that joint-parity measurements (or stabilizers in more complex devices \cite{Landau2016,Plugge2016}) can be accessed directly without ancilla qubits and initial entanglement operations which are necessary in transmon architectures \cite{Kelly2015}.

To understand how this works, we study the final reduced density matrix of the qubit system, $\rho_\mathrm{MBQ,f}$, given that it started with density matrix $\rho_\mathrm{MBQ,0}$. For the dot system initially in state $|2_d\rangle$, one finds
\begin{equation}\label{MBQredfinal}
 \rho_\mathrm{MBQ,f}=\mathrm{Tr}_\mathrm{env-meas}\,\left[U^\dagger(t)\left(\rho_\mathrm{MBQ,0}\otimes |2_d\rangle\langle2_d|\right)U(t)\right],
\end{equation}
where $U(t)$ is the time-evolution operator, and the trace runs over the measurement apparatus and environmental degrees of freedom. The Hamiltonian of the system in dot basis $\lbrace |1_d\rangle,|2_d\rangle\rbrace$ takes the form (cf. equation (4))
\begin{equation}\label{Henv}
H = H_Q + \left(\begin{array}{cc} \varepsilon +\hat Q_1& \hat{t}\\
  \hat{t}^\dagger\, & \, -\varepsilon + \hat Q_2
  \end{array}\right),
\end{equation}
where $\hat{Q}_{1,2}$ are operators that describe coupling of the charges $q_{1,2}$ on dots 1 and 2 to the environment (described by $H_Q$). For the two-qubit example mentioned above, the tunneling-amplitude is given by $\hat{t}=t_a \hat{z}_a + t_b \hat{z}_b$. At the end of the qubit readout procedure, the dot system is tuned out of resonance and the electron ends up in either $|2_d\rangle$ or $|1_d\rangle$, which can be confirmed by a measurement {as used for MBQ readout before}. We therefore consider two cases: either the electron ends up in the original state in {$|2_d\rangle$}, or in the other dot state $|1_d\rangle$.

If we find the dot state $|2_d\rangle$, the reduced density matrix is $\langle 2_d| \rho_\mathrm{MBQ,f}|2_d\rangle$. It is straightforward to see (by expansion of the time-evolution operator in powers of $\hat{t},~\hat{t}^\dagger$) that then only terms with exactly as many forward ($\hat{t}$) as backward-tunneling ($\hat{t}^\dagger$) events survive, i.e., terms that depend on $|\hat{t}|$ but not on the tunneling amplitude $\hat t$ itself. For the two-qubit example these terms depend only on $\hat{z}_a\hat{z}_b$. In other words, even though the dot system is collapsed and dephases during the readout process, only the relative phase - encoded in $|\hat t|$ - is affected. Thus, with a successful readout of the Rabi frequency $\omega_{z_az_b=\pm1}$ the desired projection of the MBQ density matrix to a coherent subspace, $z_az_b=+1$ or $z_az_b=-1$ is ensured.

On the other hand, if we find a final dot state $|1_d\rangle$ the electron has been transferred to the other dot during the readout procedure, such that there must be exactly one extra tunneling $\hat{t}= t_a\hat z_a + t_b\hat z_b$ (not balanced by back-tunneling $\hat{t}^\dagger$), and the MBQ states have acquired an additional phase factor. To recover the desired final density matrix, we can apply a confirmed electron transfer between the two dots, $1\to 2$. As discussed in the main text, such a transfer then applies the tunneling operator $\hat{t}^\dagger$,
independent of adiabaticity. After the confirmed transfer to the final state $|2_d\rangle$, we are in the same situation as before, ending up with the desired projection. Moreover, we note that after a joint-parity readout with result (say) $z_az_b = +$, because of $t^{(+,+)} = -t^{(-,-)}$, the recovery operation is identified by a relative sign between the two allowed states $|00\rangle_{ab}$ and $|11\rangle_{ab}$ in that subspace. In fact, by keeping track of the initial and final quantum dot states one can take this phase into account without physically applying the recovery operation. The above arguments can be generalized to situations with more MBQs between the quantum dots and to stabilizers in more complex systems \cite{Plugge2016}.

We have thus established that MBQ quantum information is well-protected during readout and manipulation because it is hosted in cotunneling links. This is in contrast to, e.g., spin qubits where the quantum information is transferred to quantum dot states themselves. Nevertheless, there will still be residual mechanisms for dephasing. One source could be charged two-level systems that couple to the excited charge states of the MBQ such that the fluctuator acquires which-path information during cotunneling events. This effect is, however, suppressed because the relevant time scale for tunneling is $\sim E_C^{-1}$ and therefore fluctuators with characteristic times longer than this cannot obtain significant which-path information. Moreover, near the center of the Coulomb valley where the MBQ is charge symmetric addition and removal of charges cost the same energy $E_C$, and MBQ charge fluctuations due to different cotunneling events thus tend to average out. Finally, we mention that coupling to electromagnetic fluctuations that mixes the MBS with above-gap states could lead to dephasing. This effect is suppressed by the topological gap, but a more detailed calculation is needed to determine its importance.

\end{document}